EUROPEAN SOCIETY OF RADIOLOGY
European Radiology

# Diagnostic performance of deep learning for predicting glioma isocitrate dehydrogenase and 1p/19q co-deletion in MRI: a systematic review and meta-analysis

Somayeh Farahani[1,2,3]* , Marjaneh Hejazi[1] , Mehnaz Tabassum[2,3] , Antonio Di Ieva[3] ,
Neda Mahdavifar[4] and Sidong Liu[2,3]

## Abstract

**Objectives**  We aimed to evaluate the diagnostic performance of deep learning (DL)-based radiomics models for the noninvasive prediction of isocitrate dehydrogenase (IDH) mutation and 1p/19q co-deletion status in glioma patients using MRI sequences, and to identify methodological factors influencing accuracy and generalizability.

**Materials and methods**  Following PRISMA guidelines, we systematically searched major databases (PubMed, Scopus, Embase, Web of Science, and Google Scholar) up to March 2025, screening studies that utilized DL to predict IDH and 1p/19q co-deletion status from MRI data. We assessed study quality and risk of bias using the Radiomics Quality Score and the QUADAS-2 tool. Our meta-analysis employed a bivariate model to compute pooled sensitivity and specificity, and meta-regression to assess interstudy heterogeneity.

**Results**  Among the 1517 unique publications, 104 were included in the qualitative synthesis, and 72 underwent meta-analysis. Pooled estimates for IDH prediction in test cohorts yielded a sensitivity of 0.80 (95% CI: 0.77–0.83) and specificity of 0.85 (95% CI: 0.81–0.87). For 1p/19q co-deletion, sensitivity was 0.75 (95% CI: 0.65–0.82) and specificity was 0.82 (95% CI: 0.75–0.88). Meta-regression identified the tumor segmentation method and the extent of DL integration into the radiomics pipeline as significant contributors to interstudy variability.

**Conclusion**  Although DL models demonstrate strong potential for noninvasive molecular classification of gliomas, clinical translation requires several critical steps: harmonization of multi-center MRI data using techniques such as histogram matching and DL-based style transfer; adoption of standardized and automated segmentation protocols; extensive multi-center external validation; and prospective clinical validation.

### Key Points

***Question*** *Can DL based radiomics using routine MRI noninvasively predict IDH mutation and 1p/19q co-deletion status in gliomas, and what factors affect diagnostic accuracy?*
***Findings*** *Meta-analysis showed 80% sensitivity and 85% specificity for predicting IDH mutation, and 75% sensitivity and 82% specificity for 1p/19q co-deletion status.*
***Clinical relevance*** *MRI-based DL models demonstrate clinically useful accuracy for noninvasive glioma molecular classification, but data harmonization, standardized automated segmentation, and rigorous multi-center external validation are essential for clinical adoption.*

---

*Correspondence:
Somayeh Farahani
somayeh.farahani@hdr.mq.edu.au
Full list of author information is available at the end of the article

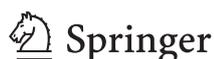

**Graphical Abstract**

## Diagnostic performance of deep learning for predicting glioma isocitrate dehydrogenase and 1p/19q co-deletion in MRI: a systematic review and meta-analysis

Can deep learning based radiomics using routine MRI noninvasively predict IDH mutation and 1p/19q co-deletion status in gliomas?

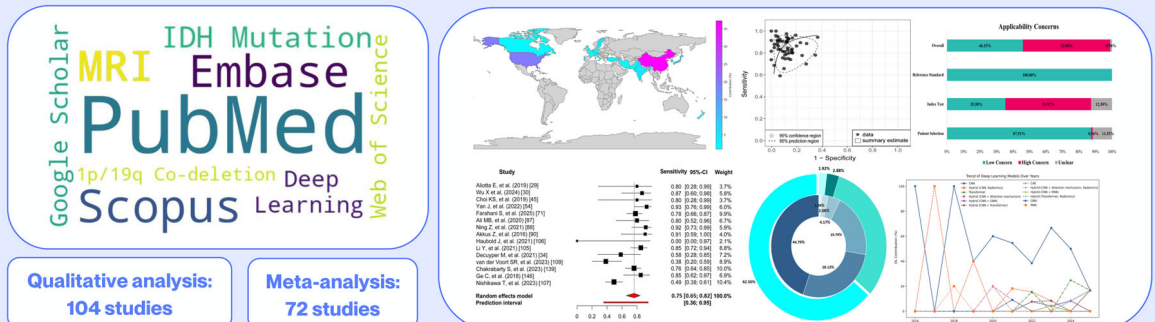

Qualitative analysis: 104 studies

Meta-analysis: 72 studies

Meta-analysis: IDH prediction 80% sensitivity, 85% specificity; 1p/19q 75% sensitivity, 82% specificity; heterogeneity driven by segmentation & deep learning integration

Eur Radiol (2025) Farahani S, Hejazi M, Tabassum M, Di Ieva A, Mahdavifar N, Liu S; DOI: 10.1007/s00330-025-11898-2

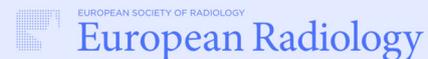

EUROPEAN SOCIETY OF RADIOLOGY
European Radiology

## Introduction

Gliomas, the most common and lethal primary central nervous system (CNS) tumors, show significant histological and molecular variability, making accurate diagnosis essential [1]. Key genetic markers, including isocitrate dehydrogenase (IDH) mutation and 1p/19q co-deletion, guide classification and treatment decisions [2]. Traditional biopsies are invasive and limited by tumor heterogeneity [3]. MRI is central to noninvasive glioma assessment, supported by the European Association of Neuro-Oncology (EANO) guidelines [4]. However, interpreting MRI data can be challenging due to human limitations and radiological "mimics," which make distinguishing gliomas from conditions such as inflammatory diseases, stroke, and infections difficult [5].

Advancements in radiomics have begun to address these challenges by extracting intricate features from medical images [6]. Radiomics analysis involves two primary methodologies: feature-engineered and deep learning (DL)-based radiomics modeling [7]. The former involves processes such as image segmentation, feature extraction, and statistical analysis, each of which significantly influences subsequent outcomes, particularly in MRI models [8]. Subjective handcrafted features spurred DL integration into radiomics. DL can replace individual steps or operate end-to-end for direct classification [7, 9].

Since the introduction of DL into radiomics, numerous studies have predicted IDH and 1p/19q co-deletion [10–12]. Given the extensive research, there is a critical need for a systematic review to synthesize and thoroughly quantify existing data. Current reviews often focus on conventional radiomics, primarily analyzing radiomic features with machine learning methods. Additionally, some works concentrate solely on specific glioma grades or particular MRI modalities (e.g., dynamic susceptibility contrast (DSC) MR perfusion imaging and T2-FLAIR mismatch) for predicting IDH mutation or 1p/19q co-deletion, often neglecting the simultaneous prediction of these biomarkers across various glioma grades and imaging techniques [13–17]. To address this gap, our study conducts a comprehensive systematic review and meta-regression to evaluate the accuracy and reliability of DL-based models in predicting IDH mutations and 1p/19q co-deletion using MRI, thereby consolidating evidence on their effectiveness.

## Methods

We performed a PRISMA-guided systematic review and meta-analysis (PROSPERO: CRD42024542505) [18].

### Search strategy and study selection

We systematically searched the PubMed, Scopus, Embase, Web of Science, and Google Scholar for DL-based radiomics studies in glioma up to March 28, 2025, with no time or language restrictions (Supplementary Section 1). We also screened relevant article bibliographies for further identification. The inclusion criteria were studies of gliomas (any World Health Organization grade) that predicted IDH and/or 1p/19q co-deletion status using MRI and incorporated DL algorithms in their radiomics workflow. For inclusion in the meta-analysis, studies had to report sufficient information to allow reconstruction of a $2 \times 2$ diagnostic table. Those without sufficient validation metrics were restricted to qualitative synthesis. Non-original and non-human studies were excluded. Records were managed via Zotero software (version 6.0.36). Two reviewers (S.F. and M.T.) independently screened the abstracts and full texts in two rounds, resolving disagreements through discussion.

### Data extraction

Two reviewers (S.F. and M.T.) independently collected data on study design, patient characteristics, datasets used, MRI sequences, data augmentation techniques, and computational methodologies using a standardized form (Supplementary Section 2). Performance metrics for constructing the diagnostic confusion matrix were obtained from both internal validation methods (e.g., k-fold and leave-one-out cross-validation) and test datasets, prioritizing external validation cohorts when available, or otherwise using held-out test sets.

When diagnostic table counts were not explicitly reported, we first contacted the corresponding authors by email. If no data were provided, we reconstructed these values from reported sensitivity, specificity, and total sample size using standard formulas. When only receiver-operating characteristic (ROC) curves were available, we extracted sensitivity and specificity at the point closest to the top-left corner (Youden index) using WebPlotDigitizer v4.7. All imputed counts were rounded to the nearest whole number. No imputation was performed for missing clinical or demographic covariates; these data were excluded from subgroup analyses. This process aligns with Cochrane Handbook guidance for transparency and reproducibility [19]. In publications reporting multiple DL models or MRI modalities, the best-performing model was selected for the meta-analysis. However, the full range of results was included and analyzed separately in the subgroup analyses.

### Quality assessment

The risk of bias and applicability concerns were evaluated via a modified QUADAS-2 tool [20], which incorporates relevant items from the Checklist for Artificial Intelligence in Medical Imaging (CLAIM) and the radiomics quality score (RQS). Key considerations included clarity in imaging protocols, appropriate data selection and missing data handling, use of reliable reference standards, and avoidance of severe genotype imbalances. Additionally, the index test evaluation assessed the use of multiple segmentations, and the robustness of the model predictions. Concerns about applicability, particularly regarding validation on external datasets, were addressed to ensure generalizability across diverse clinical settings. If the data were insufficient, we contacted the authors for clarification via email. Moreover, the methodologies, strengths, limitations, quality, and translatability of studies were evaluated using RQS, which assesses each study on 16 components, with cumulative scores ranging from $-8$ to 36 [21]. Three reviewers (S.F. and N.M. for QUADAS-2 and S.F. and M.T. for RQS) independently conducted assessments, resolving discrepancies through discussion (Supplementary, Sections 3 and 4).

### Statistical analysis

A bivariate random effects model was used to pool the sensitivity, specificity, and 95% confidence intervals (CIs) across studies ($\geq 5$) and summary receiver operating characteristic (SROC) curves. Heterogeneity was evaluated using Cochran's $Q$-test, the $I^2$ statistic, prediction intervals ($p < 0.05$), and the Spearman correlation coefficients (SCC) between sensitivity and the false positive rate (threshold effect indicated by an $SCC > 0.6$) [22, 23]. Subgroup analyses explored sources of heterogeneity by subgroup analyses in instances with enough studies [24]. A leave-one-out meta-analysis assessed each study's impact on effect size. Publication bias was evaluated using funnel plots and Egger's test. The Trim and Fill method by Duval and Tweedie was applied to adjust pooled sensitivity and specificity estimates in the presence of asymmetry. Statistical power was also calculated across effect sizes [25]. Analyses were performed with R packages 'mada', 'metameta', 'metafor' (R v4.4.1), and Meta-BayesDTA (v1.5.2) [26].

## Results

### Study characteristics

A total of 1517 unique publications were initially identified through primary searches and relevant study bibliographies. Following screening and full-text reviews, 104 studies were eligible for qualitative analysis, of which 72 were included in the meta-analysis (Fig. 1). One study [27] was excluded because it served solely for the external validation of another study [28]. Details on the inclusion and exclusion status of studies in the meta-analysis are summarized in Supplementary Table 4.

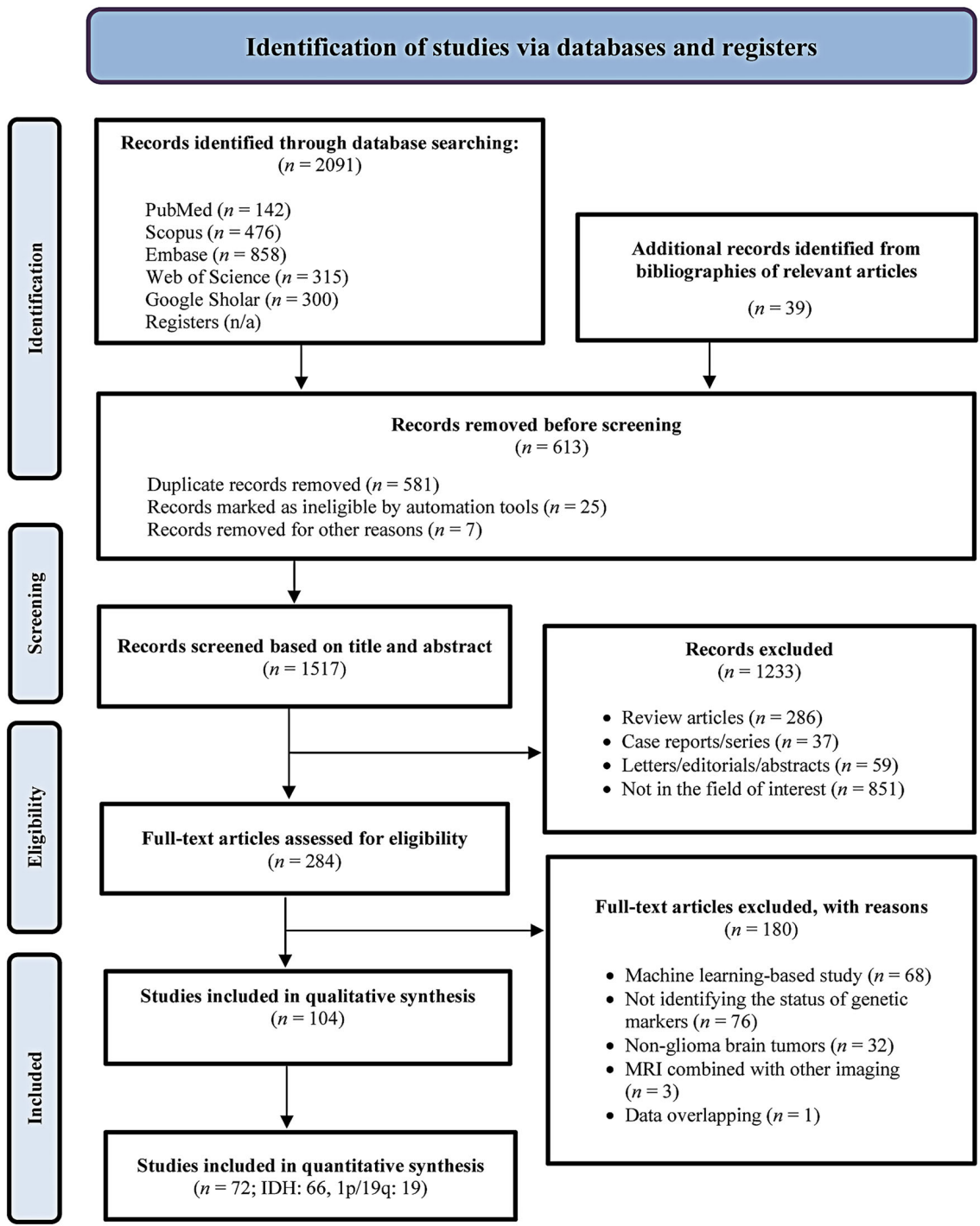

**Fig. 1** Flow diagram of the study selection process

Our analysis revealed that China and the USA dominate global research in this field, significantly outpacing other countries' publication volume (Fig. 2A). Additionally, the surveyed studies spanned various sample sizes, ranging from 41 [29] to 2776 patients [30]. Over 23% of the included studies employed genotyping methods, such as immunohistochemistry, DNA sequencing, or fluorescence

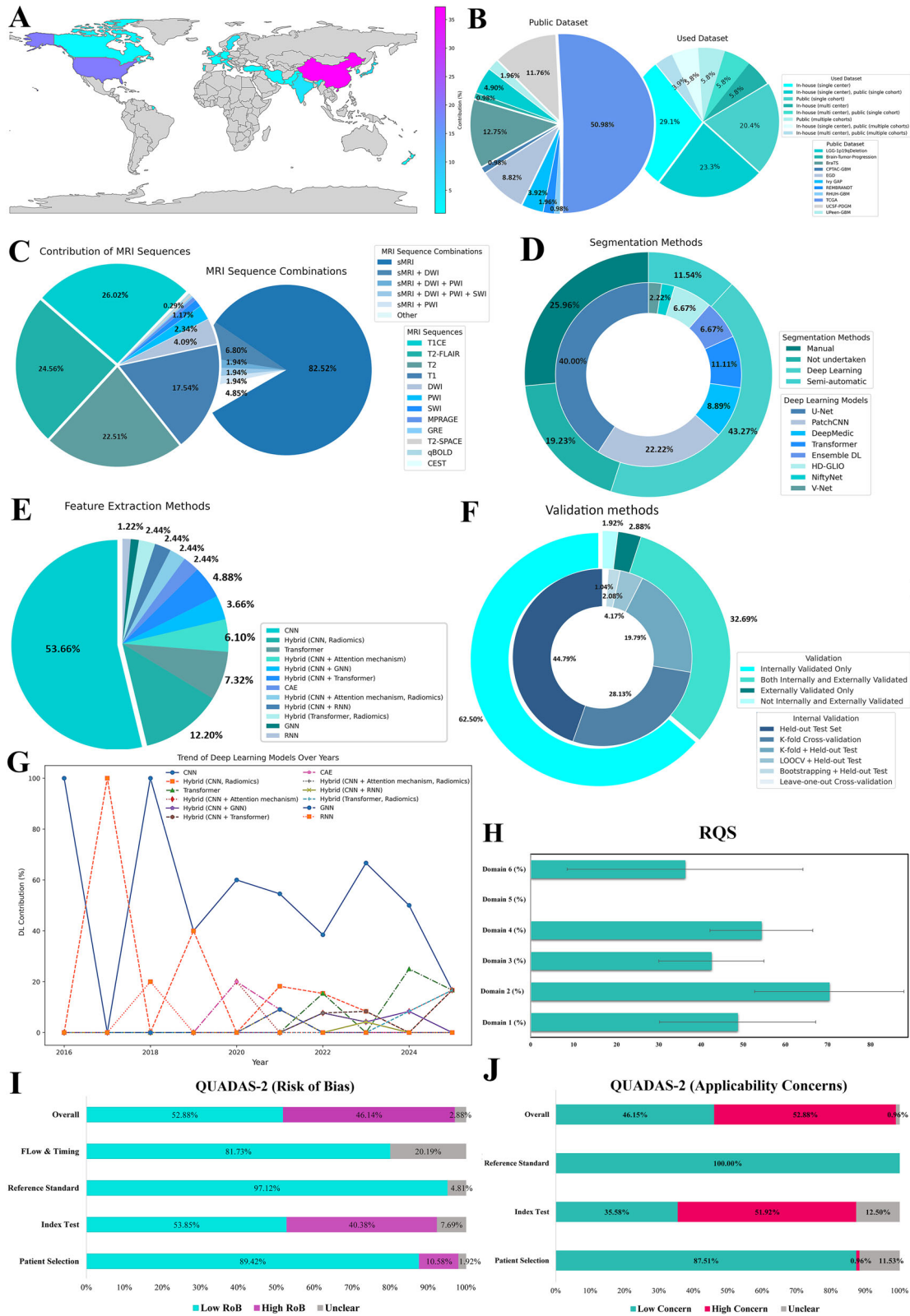

**Fig. 2** (See legend on next page.)

(see figure on previous page)

**Fig. 2 A** A Heatmap depicting global publication trends in MRI-based DL models for predicting IDH and 1p/19q codeletion status. **B** Distribution of dataset types in our included studies, highlighting the prevalence of public, in-house, and combined data sources. **C** Distribution of MRI sequences used in our included studies, highlighting the diversity and frequency of sequence combinations employed in research. **D** Breakdown of segmentation methods used in 104 studies. The outer ring displays the overall segmentation techniques, whereas the inner donut highlights the distribution of DL methods. **E** Distribution of DL architectures used for feature extraction, highlighting the predominance of CNN-based models. **F** Validation methods employed in the studies. The outer ring shows the overall approach to validation, while the inner ring details specific internal validation methods. **G** Temporal trends (2016–2025) in DL model choice in feature extraction, illustrating the shift from CNN-only workflows to autoencoders, transformers, and other hybrids. **H** Mean RQS ± SE across six quality domains (D1 (protocol quality), D2 (feature selection and validation), D3 (biologic/clinical validation and utility), D4 (model performance index), D5 (level of evidence), and D6 (open science and data). **I** QUADAS-2 risk-of-bias profile for the included studies. **J** QUADAS-2 applicability concerns, with high concern in 55 studies. IDH, isocitrate dehydrogenase; CEST, chemical-exchange saturation transfer; DWI, diffusion-weighted imaging; DSC-PWI, dynamic susceptibility-contrast perfusion-weighted imaging; MRAGE, magnetization-prepared rapid acquisition gradient-echo; qBOLD, quantitative blood-oxygen-level-dependent; sMRI, structural MRI; SWI, susceptibility-weighted imaging; TR-SPACE, T2 sampling perfection with application-optimized contrasts using different flip-angle evolutions; DL, deep learning; AE, autoencoder; CNN, convolutional neural network; GNN, graph neural network; nnUNet, no-new-net U-Net framework; RNN, recurrent neural network; VNet, volumetric neural network; LOOCV, leave-one-out cross-validation; ROI, region of interest; QUADAS-2, quality assessment of diagnostic accuracy studies, version 2; LoB, risk of bias; RQS, radiomics quality score; D, domain; TCGA, the Cancer Genome Atlas; Ivy GAP, Ivy Glioblastoma Atlas Project; RHUH-GBM, Río Hortega University Hospital Glioblastoma Dataset; UPenn-GBM, University of Pennsylvania Glioblastoma Dataset; UCSF-PDGM, University of California, San Francisco Preoperative Diffuse Glioma MRI; EGD, Erasmus Glioma Database; LGG-1p19qDeletion, LGG-1p19q deletion dataset

in situ hybridization, to directly assess genetic variants. In contrast, 44% used molecular typing, which classifies tumors into clinically relevant subgroups based on their genotyping results [2]. Approximately 20% of studies combined both approaches by integrating multiple cohorts, whereas 12% did not report the reference standard used to determine biomarker status (Table 1).

In our qualitative analysis, we identified three primary imaging data sources: private (in-house), public, and a combination of both (Fig. 2B). In-house collections accounted for 35% of the data, whereas 26% relied solely on public datasets, especially The Cancer Imaging Archive (TCIA). Approximately 38% of the studies combined both sources for enhanced research robustness and data diversity. Furthermore, around 63% of the studies implemented data augmentation techniques—either conventional methods [12, 31–35] or generative adversarial networks (GANs) [36–38]—to mitigate overfitting and address class imbalance related to genotype distributions (Supplementary Table 5).

The extensive use of public datasets influenced MRI sequence choices, with conventional methods employed in 81% of the studies. The combination of T1, T1CE, T2, and T2-FLAIR sequences was most prevalent, accounting for 40% of cases. Notably, T1CE was the most frequently utilized, appearing in 26% of studies, followed by T2-FLAIR and T2. Advanced imaging techniques, such as diffusion- and perfusion-weighted imaging, were less commonly employed, appearing in 8% [29, 33, 39–44] and 3% [45–47] of studies, respectively, or in 5% [48–50] when used in combination (Fig. 2C).

DL methods, predominantly based on convolutional neural networks (CNNs), were used for tumor segmentation in 43% of the studies, followed by manual segmentation (26%) and semi-automatic methods (12%) (Fig. 2D). Twenty studies did not incorporate precisely delineated tumors into their prediction models; among these, 55% relied on whole preprocessed MRI slices, 30% utilized regions of interest (ROIs) confined to bounding boxes, and 15% used cropped tumor-bearing regions. Feature extraction primarily relied on CNN-based models, such as ResNet [35, 43, 51–57], DenseNet [32, 58–60], and Inception [61, 62], which were employed in more than half of the studies. Hybrid CNN–radiomics approaches [63–67] and transformer-based models [68–75] followed, appearing in approximately 12% and 7% of cases, respectively. Less commonly used methods included hybrid DL models [36, 41, 64, 65, 76–86], CAE [87, 88], graph neural networks (GNNs) [89], and recurrent neural networks (RNNs) [45] (Fig. 2E). More details on DL model architectures are provided in Supplementary Table 5.

Our review highlights the rise in DL-based radiomics research since 2016 [90]. Initially, CNNs dominated exclusively, making up 100% of methodologies from 2016 to 2018. However, diversification has since increased. By 2020, CNNs still led at 60%, with CAEs [87] and hybrid models combining CNNs with attention mechanisms [36] emerging as alternative models. Transformers, introduced in recent years, peaked at 25% in 2024 (Fig. 2G), indicating a shift to more complex architectures. These networks were primarily integrated into prediction frameworks in an end-to-end manner [33, 41, 51, 52, 62, 71–73,75, 77, 82, 84, 87, 91–94]. In some cases, DL was specifically applied for tumor segmentation [95–97], image preprocessing [98], or classification [48, 99] within the radiomics pipeline.

Pretrained models were used in approximately 38% of the studies (Supplementary Table 5). Among these, 75% fine-tuned the pretrained weights on their own training data, while the remaining studies applied them "as-is," primarily for tumor segmentation. Notably, several

**Table 1** Characteristics of the 104 included studies, all of which employed a retrospective design

| Study | Total no. pts | Genes | Grade | Reference standard | Dataset(s) | MRI | Segmentation | Feature extraction | Validation | AUC |
|---|---|---|---|---|---|---|---|---|---|---|
| Choi et al [45] | 463 | IDH, 1p/19q | 2, 3, 4 | IDH: IHC, 1p/19q: FISH | In-house (single center) | T1, T1CE, T2, T2-FLAIR, DSC | CNNs | RNNs | Internally validated only | IDH: 95%, 1p/19q: 89% |
| Fukuma et al [31] | 164 | IDH | 2, 3 | IDH1/2: Sanger sequencing, pyrosequencing | In-house (multi-center) | T1, T1CE, T2, T2-FLAIR | Manual | Hybrid (CNNs, Radiomics) | Internally validated only | IDH: 69.9% |
| Ge et al [36] | 167 | IDH | 2, 3, 4 | TCGA | Public (TCGA) | T1, T1CE, T2, T2-FLAIR | Manual | CNN with attention-weighted feature fusion | Internally validated only | IDH: 88.82%* |
| Li et al [10] | 151 | IDH | 2 | IDH1: Sanger sequencing | In-house (single center) | T1CE, T2-FLAIR | CNNs | Hybrid (CNNs, Radiomics) | Internally validated only | IDH: 96.15% |
| Liang et al [32] | 167 | IDH | 2, 3, 4 | TCGA | Public (TCGA) | T1, T1CE, T2, T2-FLAIR | Manual | 3D DenseNet | Internally validated only | IDH: 85.7% |
| Wu et al [129] | 105 | IDH | NR | NR | In-house (single center) | T1CE, T2-FLAIR | CNNs | Radiomics | Internally validated only | IDH: 88% |
| Kim et al [12] | 143 | 1p/19q | 2, 3, 4 | BraTS 2017 | Public (BraTS 2017) | T1, T1CE, T2-FLAIR | Manual | Hybrid (CNNs, Radiomics) | Internally validated only | 1p/19q: 69.1% |
| Chang et al [11] | 259 | IDH, 1p/19q | 2, 3, 4 | TCGA | Public (TCGA) | T1, T1CE, T2, T2-FLAIR | Manual | CNNs | Internally validated only | IDH: 91%, 1p/19q: 88% |
| Ali et al [87] | 161 | IDH, 1p/19q | 2 | NR | In-house (multi-center) | T1CE, T2-FLAIR | Not undertaken (whole MRI slices) | Multi-stream convolutional autoencoder (CAE) | Internally validated only | IDH: 81.19%, 1p/19q: 74.81%* |
| Tang et al [33] | 93 | IDH, 1p/19q | 4 | Genomic sequencing | In-house (single center) | T1CE, DTI | Radiomics (manual); DL (not undertaken, bounding box) | CNNs | Internally validated only | IDH: 94.6%, 1p/19q: 88.1%* |
| Decuyper et al [34] | 466 | IDH, 1p/19q | 2, 3, 4 | TCGA, In-house dataset: IDH: IHC, 1p19q: FISH | In-house (single center), public (TCGA, LGG-1p19qDeletion, BraTS 2019) | T1, T1CE, T2, T2-FLAIR | U-Net | CNNs | Both internally and externally | IDH: 86.23%, 1p/19q: 86.61% |

**Table 1** continued

| Study | Total no. pts | Genes | Grade | Reference standard | Dataset(s) | MRI | Segmentation | Feature extraction | Validation | AUC |
|---|---|---|---|---|---|---|---|---|---|---|
| Ning et al [88] | 645 | IDH, 1p/19q | 2, 3, 4 | TCGA, In-house dataset: NR | In-house (single center), public (TCGA) | T1CE, T2-FLAIR | Manual | CAE | Externally validated only | IDH: 92.4%, 1p/19q: 90.2% |
| van der Voort et al [130] | 1748 | IDH, 1p/19q | 2, 3, 4 | REMBRANDT, CPTAC-GBM, Ivy GAP, BraTS, Brain-Tumor-Progression, In-house dataset: NR | In-house (multi-center), public (REMBRANDT, CPTAC-GBM, Ivy GAP, BraTS, Brain-Tumor-Progression) | T1, T1CE, T2, T2-FLAIR | CNNs | Multi-task CNNs | Both internally and externally | IDH: 90%, 1p/19q: 85% |
| Cluceru et al [42] | 531 | IDH, 1p/19q | 2, 3, 4 | TCGA, UCSF | Public (TCGA, UCSF) | T1CE, T2, T2-FLAIR, ADC | Manual | CNNs | Both internally and externally | Overall: 85.70%* |
| Haubold et al [106] | 145 | IDH, 1p/19q | 2, 3, 4 | NR | In-house (single center) | T1, T1CE, T2-FLAIR | DeepMedic | Radiomics | Internally validated only | IDH: 86.1%, 1p/19q: 71.1% |
| Tupe-Waghmare et al [35] | 269 | IDH, 1p/19q | 4 | TCGA, In-house dataset: NR | In-house (single center), public (TCGA) | T1CE, T2, T2-FLAIR | DeepMedic | ResNet50 | Internally validated only | IDH: 80.04%, 1p/19q: 88.73%* |
| Matsui et al [52] | 217 | IDH, 1p/19q | 2, 3 | IDH: IHC, 1p19q: FISH | In-house (single center) | T1, T2, T2-FLAIR | Not undertaken (bounding box) | ResNet | Internally validated only | NR |
| Chang et al [51] | 496 | IDH | 2, 3, 4 | TCGA, In-house dataset: IHC, next-generation sequencing, mass spectrometry-based mutation genotyping (OncoMap), capture-based sequencing (OncoPanel) | In-house (multi-center), public (TCGA) | T1, T1CE, T2, T2-FLAIR | Not undertaken (bounding box) | ResNet | Internally validated only | IDH: 91% |
| Calabrese et al [49] | 256 | IDH | 4 | TCGA, Genetic sequencing | In-house (single center), public (TCGA-GBM) | T1, T1CE, T2, T2-FLAIR, SWI, DWI, ASL, DTI | CNNs | Radiomics | Both internally and externally validated | IDH: 63% |

**Table 1** continued

| Study | Total no. pts | Genes | Grade | Reference standard | Dataset(s) | MRI | Segmentation | Feature extraction | Validation | AUC |
|---|---|---|---|---|---|---|---|---|---|---|
| Ai et al [77] | 235 | IDH | 2, 3, 4 | BraTS 2019, TCGA | Public (BraTS-2019, TCGA) | T2-FLAIR | TDABNet | TDABNet | Internally validated only | IDH: 96.44% |
| Chaddad et al [101] | 83 | IDH, 1p/19q | 2, 3 | TCGA | Public (TCGA) | T1, T1CE, T2, T2-FLAIR | Semi-automatic | CNNs | Internally validated only | IDH: 70.0% |
| Chakrabarty et al [131] | 1047 | IDH, 1p/19q | 2, 3, 4 | TCGA, BraTS, Ivy GAP, EGD, LGG-1p19qDeletion, In-house dataset: IDH: IHC, 1p19q: FISH | In-house (single center), public (TCGA, BraTS, Ivy GAP, EGD, LGG-1p19qDeletion) | T1CE, T2, T2-FLAIR | CNNs | CNNs | Both internally and externally validated | IDH: 93.3%, 1p/19q: 84.2% |
| Chakrabarty et al [91] | 546 | IDH | 2, 3, 4 | TCGA, BraTS, Ivy GAP, In-house dataset: NR | In-house (single center), public (TCGA, BraTS, Ivy GAP) | T1CE, T2, T2-FLAIR | 3D Mask R-CNN | 3D Mask R-CNN | Both internally and externally validated | IDH: 87.1% |
| Chen et al [82] | 271 | IDH, 1p/19q | NR | BraTS, In-house dataset: NR | In-house (single center), public (BraTS) | T1, T1CE, T2, T2-FLAIR | Not undertaken (bounding box) | WSOFNet | Both internally and externally validated | IDH: 96.55% |
| Chu et al [92] | 190 | IDH | 2, 3, 4 | BraTS 2019 | Public (BraTS 2019) | T1, T1CE, T2, T2-FLAIR | UNet++ | UNet++ | Internally validated only | IDH: 95.22%* |
| Buz-Yaluğ et al [46] | 162 | IDH | 4 | Minisequencing | In-house (single center) | T1, T1CE, T2, DSC | Manual | ResNet50, VGG16 | Internally validated only | IDH: 89% |
| Calabrese et al [50] | 400 | IDH | 4 | UCSF | Public (UCSF) | T1, T1CE, T2, T2-FLAIR, SWI, ASL, DWI, HARDI | CNNs | Hybrid (CNNs, Radiomics) | Internally validated only | IDH: 96% |
| Cheng et al [79] | 218 | IDH | 2, 3, 4 | BraTS 2020 | Public (BraTS 2020) | T1, T1CE, T2, T2-FLAIR | Manual | Hybrid CNN-Transformer encoder | Internally validated only | IDH: 90.37% |

**Table 1** continued

| Study | Total no. pts | Genes | Grade | Reference standard | Dataset(s) | MRI | Segmentation | Feature extraction | Validation | AUC |
|---|---|---|---|---|---|---|---|---|---|---|
| Choi et al [95] | 136 | IDH | 4 | TCGA, In-house dataset: Sanger sequencing | In-house (single center), public (TCGA) | T2 | V-Net | Radiomics | Externally validated only | IDH: 85.7% |
| Choi et al [28] | 1166 | IDH, 1p/19q | 2, 3, 4 | TCGA, In-house dataset: NR | In-house (multi-center), public (TCGA) | T1CE, T2, T2-FLAIR | U-Net | Hybrid (CNNs, Radiomics) | Both internally and externally validated | IDH: 94% |
| Gore et al [132] | 217 | IDH | 2, 3, 4 | TCGA | Public (TCGA) | T1, T1CE, T2, T2-FLAIR | Not undertaken (cropped tumor-bearing regions) | CNNs | Internally validated only | IDH: 93.67%* |
| Karami et al [43] | 146 | IDH, 1p/19q | 2, 3, 4 | NR | In-house (single center) | T1CE, T2, T2-FLAIR, multi-shell diffusion | Semi-automatic (HD-GLIO) | ResNet10 | Internally validated only | IDH: 75%, 1p/19q: 72%* |
| Liu et al [98] | 78 | IDH | 4 | NR | In-house (single center) | T1, T1CE, T2, T2-FLAIR, DSC, DWI | nnU-Net | Radiomics (quantitative radiological parameters) | Not internally and externally validated | IDH: 81.5% |
| McHugh et al [59] | 1158 | IDH, 1p/19q | 2, 3, 4 | TCGA, EGD, In-house dataset: IDH: IHC, genetic sequencing; 1p19q: FISH | In-house (single center), public (TCGA, EGD) | T1CE, T2, T2-FLAIR | Dense U-Net | 2D Dense U-Nets | Both internally and externally validated | IDH: 95.8%, 1p/19q: 85.4% |
| Moon et al [53] | 878 | IDH | 2, 3, 4 | TCGA, In-house dataset: NR | In-house (single center), public (TCGA) | T1CE, T2-FLAIR | U-Net | ResNet50 | Both internally and externally validated | IDH: 83.3% |
| Nalawade et al [62] | 260 | IDH | 2, 3, 4 | TCGA | Public (TCGA) | T2 | Not undertaken (whole MRI slices) | DenseNet-161, ResNet-50, Inception-v4 | Internally validated only | IDH: 84% |

**Table 1** continued

| Study | Total no. pts | Genes | Grade | Reference standard | Dataset(s) | MRI | Segmentation | Feature extraction | Validation | AUC |
|-------|------|-------|-------|-------------------|-----------|-----|--------------|-------------------|-----------|-----|
| Nalawade et al [60] | 368 | IDH, 1p/19q | 2, 3, 4 | TCGA | Public (TCGA) | T2 | Dense U-Net | 3D Dense-Unet | Internally validated only | IDH: 68%, 1p/19q: 82%* |
| Pasquini et al [133] | 100 | IDH | 4 | IHC, Sanger sequencing | In-house (single center) | T1, T2, T2-FLAIR, MPRAGE, rCBV, ADC | Not undertaken (Bounding box) | CNNs | Internally validated only | IDH: 86%* |
| Rui et al [103] | 42 | IDH | 2, 3, 4 | IHC | In-house (single center) | T1CE, T2-FLAIR, QSM | Semi-automatic (ITK-SNAP) | CNNs | Internally validated only | IDH: 89%* |
| Safari et al [134] | 105 | IDH | 2, 3 | TCGA | Public (TCGA) | T1, T1CE, T2, T2-FLAIR | Semi-automatic | Shuffle-ResNet | Internally validated only | IDH: 94.3% |
| Zhang et al [84] | 759 | IDH | 2, 3, 4 | TCGA, In-house dataset: IHC, Genomic sequencing | In-house (multi-center), public (TCGA) | T1, T1CE, T2, T2-FLAIR | nnU-Net | MFEFnet | Both internally and externally validated | IDH: 85.64% |
| Zhang et al [66] | 486 | IDH | 2, 3, 4 | TCGA, In-house dataset: IHC, Sanger sequencing | In-house (single center), public (TCGA) | T1, T1CE, T2, T2-FLAIR | nnU-Net | Radiomics | Internally validated only | IDH: 92% |
| Yogananda et al [135] | 1849 | IDH | NR | TCIA, EGD, UCSF, In-house dataset: Sanger sequencing, next-generation sequencing, IHC | In-house (multi-center), public (TCGA, EGD, UCSF) | T1, T1CE, T2, T2-FLAIR | FeTS tool, nnU-Net | nnU-Net | Both internally and externally validated | IDH: 96.46% |
| Zeng et al [67] | 110 | IDH | 2, 3, 4 | BraTS 2019, TCGA, In-house dataset: NR | In-house (single center), public (BraTS 2019) | T1, T1CE, T2, T2-FLAIR | MIDAS framework | Hybrid (CNNs, Radiomics) | Internally validated only | IDH: 86% |
| Xu et al [72] | 188 | IDH | 2, 3, 4 | IHC | In-house (single center) | T1CE, T2 | Not undertaken (whole MRI slices) | Vision Transformer (ViT) | Internally validated only | IDH: 98.2% |
| Wu et al [70] | 493 | IDH | 2, 3, 4 | TCGA, In-house dataset: IHC | In-house (single center), public (TCGA) | T2 | Manual + Bounding box | Swin Transformer, ResNet-101 | Both internally and externally validated | IDH: 87.8% |

**Table 1** continued

| Study | Total no. pts | Genes | Grade | Reference standard | Dataset(s) | MRI | Segmentation | Feature extraction | Validation | AUC |
|---|---|---|---|---|---|---|---|---|---|---|
| Wei et al [89] | 372 | IDH | NR | TCGA, Ivy-GAP, In-house dataset: NR | In-house (single center, public (TCGA, Ivy-GAP) | T1, T1CE, T2, T2-FLAIR | Manual | Graph Neural Network (GNN) | Internally validated only | IDH: 86.6%* |
| Wang et al [76] | 121 | IDH | 2, 3, 4 | TCGA, BRATS 2020 | Public (TCGA, BraTS 2020) | T1, T1CE, T2, T2-FLAIR | 3D U-Net | SGPNet | Internally validated only | IDH: 94.9% |
| Wei et al [78] | 372 | IDH | NR | TCGA, Ivy-GAP, In-house dataset: NR | In-house (single center), public (TCGA, Ivy-GAP) | T1, T1CE, T2, T2-FLAIR | Semi-automatic | Hybrid (CNNs, GNN) | Internally validated only | IDH: 89.2% |
| Wei et al [81] | 387 | IDH | NR | TCGA, In-house dataset NR | In-house (single center), public (TCGA) | T1, T1CE, T2, T2-FLAIR | nnU-Net | Hybrid (CNNs, GNN) | Internally validated only | IDH: 90.5% |
| Tripathi et al [136] | 377 | IDH, 1p/19q | 2, 3, 4 | TCGA-LGG, TCGA-GBM, LGG-1p19qDeletion | Public (TCGA-LGG, TCGA-GBM, TCGA-1p19qdeletion) | T1, T1CE, T2, T2-FLAIR | CNNs | CNNs | Internally validated only | IDH: 91.96%, 1p/19q: 87.88%* |
| Shi et al [65] | 489 | IDH | NR | Sanger sequencing | In-house (single center) | T1, T1CE, T2, T2-FLAIR | Manual | Hybrid (SA-Net (self-attention network), Radiomics) | Internally validated only | IDH: 81% |
| Shi et al [83] | 218 | IDH | NR | BraTS 2020 | Public (BraTS 2020) | T1, T1CE, T2, T2-FLAIR | TransBTS | 3D U-Net encoder with transformer (TransBTS) | Internally validated only | IDH: 90.3% |
| Yan et al [54] | 555 | IDH, 1p/19q | 2, 3 | TCGA, In-house dataset: IDH: IHC, 1p19q: FISH | In-house (single center), public (TCGA) | T1, T1CE, T2, T2-FLAIR | Manual | Deep CNN (ResNet-34 based) | Both internally and externally validated | 1p/19q: 98.3% |
| Kihira et al [104] | 239 | IDH | 2, 3, 4 | IHC | In-house (multi-center) | T2-FLAIR | U-Net | Radiomics | Both internally and externally validated | IDH: 93% |
| Sohn et al [137] | 418 | IDH | 4 | Next-generation sequencing | In-house (single center) | T1, T1CE, T2, T2-FLAIR | Semi-automatic (HD-GLIO) | Radiomics | Internally validated only | IDH: 96.4% |

**Table 1** continued

| Study | Total no. pts | Genes | Grade | Reference standard | Dataset(s) | MRI | Segmentation | Feature extraction | Validation | AUC |
|---|---|---|---|---|---|---|---|---|---|---|
| Buda et al [138] | 110 | IDH, 1p/19q | 2, 3 | TCGA | Public (TCGA) | T1, T1CE, T2, T2-FLAIR | U-Net | Radiomics | Internally validated only | NR |
| Ali et al [80] | 167 | IDH | 2, 3, 4 | TCGA, In-house dataset: NR | In-house (single center), public (TCGA) | T1CE, T2-FLAIR | Manual | EfFedDyn with an attention-weighted fusion layer | Internally validated only | IDH: 85.46%* |
| Chen et al [73] | 1153 | IDH, 1p/19q | NR | TCGA, In-house dataset: NR | In-house (single center), public (TCGA) | T1, T1CE, T2, T2-FLAIR | Not undertaken (whole MRI slices) | Vision Transformer | Both internally and externally validated | NR |
| Elyassirad et al [55] | 495 | IDH | 2, 3, 4 | UCSF | Public (UCSF) | T1, T1CE, T2-FLAIR | Manual | ResNet | Internally validated only | IDH: 90.96% |
| Fayyaz et al [93] | 89 | IDH | 2, 3, 4 | TCGA | Public (TCGA) | NR | Not undertaken (whole MRI slices) | CNNs (Xception, ResNet152V2, InceptionV3, InceptionResNetV2, NASNetLarge) | Internally validated only | IDH: 99% |
| Ge et al [139] | 159 | 1p/19q | 2 | NR | In-house (single center) | T1CE, T2 | Not undertaken (mask-enhanced whole MRI slices) | CNNs | Internally validated only | 1p/19q: 89.39% |
| Gómez Vecchio et al [56] | 469 | IDH | 2, 3 | EGD, In-house dataset: NR | In-house (multi-center), public (EGD) | T1CE, T2, T2-FLAIR | Semi-automatic | ResNet152 | Both internally and externally validated | NR |
| Hosseini et al [37] | 57 | IDH | 4 | IHC, RealTime PCR | In-house (single center) | T1CE, T2-FLAIR | Semi-automatic | Radiomics | Internally validated only | IDH: 92% |
| Jeon et al [96] | 218 | IDH, 1p/19q | 2, 3 | NR | In-house (single center) | T1, T1CE, T2, T2-FLAIR | Semi-automatic (HD-GLIO) | Radiomics (quantitative radiological parameters) | Not internally and externally validated | IDH: 69% |

**Table 1** continued

| Study | Total no. pts | Genes | Grade | Reference standard | Dataset(s) | MRI | Segmentation | Feature extraction | Validation | AUC |
|---|---|---|---|---|---|---|---|---|---|---|
| Jian et al [140] | 418 | IDH | 2, 3, 4 | UCSF | Public (UCSF-PDGM) | T1, T1CE, T2, T2-FLAIR | BraTS-based ensemble model | Radiomics | Internally validated only | IDH: 80.18% |
| Li et al [63] | 402 | IDH | 2, 3, 4 | NR | In-house (multi-center) | T2 | Manual | Hybrid (ResNet101, Radiomics) | Both internally and externally validated | IDH: 98% |
| Li et al [57] | 263 | IDH | 2, 3, 4 | IHC, Genome sequencing | In-house (single center) | T1, T1CE, T2, T2-FLAIR | Manual | ResNet50 | Internally validated only | IDH: 87.1% |
| Li et al [105] | 1016 | IDH, 1p/19q | 2, 3, 4 | IDH: IHC, Pyrosequencing; 1p/19q: FISH | In-house (single center) | T1, T1CE, T2 | Manual | Hybrid (ResNet18, Radiomics) | Internally validated only | IDH: 89%, 1p/19q: 85% |
| Lost et al [141] | 584 | IDH | 2, 3, 4 | IHC ± Sanger sequencing | In-house (multi-center) | T1CE, T2-FLAIR, PGSE, GRE | U-Net Transformer (UNETR) | Radiomics | Both internally and externally validated | IDH: 83.5% |
| Nishikawa et al [107] | 460 | IDH, 1p/19q | 2, 3, 4 | TCGA, In-house dataset: exome sequencing; Sanger sequencing; 1p/19q multiplex ligation-dependent probe amplification (MLPA) | In-house (single center), public (TCGA) | T2 | Manual | CNNs | Both internally and externally validated | IDH: 70.3%, 1p/19q: 65.1%* |
| Park et al [38] | 162 | IDH | 2, 3 | IHC, pyrosequencing | In-house (single center) | T1CE, T2-FLAIR | Not undertaken (whole MRI slices) | Radiomics (radiological parameters) | Internally validated only | IDH: 82.1% |
| Sadi-Bilmez et al [47] | 225 | IDH | 2, 3, 4 | Minisequencing, Sanger sequencing | In-house (single center) | T1, T1CE, T2, 1H-MRS, DSC | Manual | CNNs | Internally validated only | IDH: 87.8%* |
| Sairam et al [142] | 58 | IDH | 3, 4 | TCGA | Public (TCGA) | T1, T2, T2-FLAIR | Not undertaken (cropped tumor-bearing regions) | CNNs | Internally validated only | IDH: 99.1% |

**Table 1** continued

| Study | Total no. pts | Genes | Grade | Reference standard | Dataset(s) | MRI | Segmentation | Feature extraction | Validation | AUC |
|---|---|---|---|---|---|---|---|---|---|---|
| Santinha et al [143] | 231 | IDH | 2, 3, 4 | TCGA, In-house dataset: NR | In-house (multi-center), public (TCGA) | T1, T1CE, T2, T2-FLAIR | Automatic (HD-GLIO) | Radiomics | Both internally and externally validated | IDH: 83.6% |
| Shi et al [64] | 488 | IDH | NR | Sanger sequencing | In-house (single center) | T1, T2, T1CE, T2-FLAIR | Manual | Hybrid (SA-Net, Radiomics) | Internally validated only | IDH: 82% |
| Stadlbauer et al [48] | 215 | IDH | 2, 3, 4 | NR | In-house (multi-center) | T1CE, T2-FLAIR, DWI, GE-DSC, multi-echo GE & SE (qBOLD), SE-DSC, GESE-DSC | Semi-automatic | Radiomics | Both internally and externally validated | IDH: 87.9% |
| Taha et al [99] | 326 | IDH | 4 | TCGA, In-house dataset: next-generation sequencing, IHC | In-house (single center), public (TCGA) | T1CE | Semi-automatic | Radiomics | Both internally and externally validated | NR |
| Usuzaki et al [69] | 597 | IDH | 2, 3, 4 | UCSF-PDGM, UPenn-GBM | Public (UCSF-PDGM, UPenn-GBM) | T1CE | BraTS challenge model | Hybrid (vViT, Radiomics) | Both internally and externally validated | IDH: 88.7% |
| Wang et al [97] | 627 | IDH | 2, 3, 4 | TCGA, In-house dataset: IHC | In-house (single center), public (TCGA) | T1, T1CE, T2, T2-FLAIR | nnU-Net | Radiomics | Both internally and externally validated | IDH: 86% |
| Wankhede et al [144] | 253 | IDH, 1p/19q | 2 | NR | NR | T1CE, T2-FLAIR | Not undertaken (whole MRI slices) | CNNs | Internally validated only | NR |

**Table 1** continued

| Study | Total no. pts | Genes | Grade | Reference standard | Dataset(s) | MRI | Segmentation | Feature extraction | Validation | AUC |
|---|---|---|---|---|---|---|---|---|---|---|
| Yang et al [86] | 811 | IDH, 1p/19q | 2, 3, 4 | TCGA-LGG, TCGA-GBM, UCSF-PDGM, EGD. In-house dataset: NR | In-house (single center), public (TCGA-LGG, TCGA-GBM, UCSF-PDGM, EGD) | T1, T1CE, T2, T2-FLAIR | Manual | Hybrid (CNNs, Swin Transformer) | Both internally and externally validated | IDH: 73.6% |
| Yu et al [74] | 664 | IDH | 2, 3, 4 | TCGA, In-house dataset: CRISPR/Cas12a-based assay, droplet digital PCR (ddPCR), and Sanger sequencing | In-house (multi-center), public (TCGA) | T1, T1CE, T2, T2-FLAIR | Manual | Vision Transformer (ViT) | Both internally and externally validated | IDH: 89% |
| Yuan et al [40] | 226 | IDH | 2, 3, 4 | IDH: Sanger sequencing; 1p/19q: qPCR assay | In-house (single center) | T1, T2, T2-FLAIR, DTI | Semi-automatic | VGG16 | Internally validated only | IDH: 83.4% |
| Yuan et al [145] | 84 | IDH | 1, 2, 3, 4 | Next-generation sequencing, IHC | In-house (single center) | CEST, MPRAGE, T2-SPACE | Manual | Hybrid (CNNs, Radiomics) | Internally validated only | IDH: 88.68% |
| Zhang et al [100] | 162 | IDH | 2, 3, 4 | NR | In-house (single center) | T1, T1CE, T2, T2-FLAIR | Automatic (NiftyNet) | Radiomics | Internally validated only | IDH + MGMT: 95% |
| Zhang et al [146] | 502 | IDH | NR | NR | In-house (single center) | T1, T1CE, T2, T2-FLAIR | Manual | SE-Net | Internally validated only | IDH: 75.25%* |
| Zhao et al [147] | 150 | 1p/19q | 2, 3, 4 | TCGA, In-house dataset: NR | In-house (single center), public (TCGA) | T1, T1CE, T2, T2-FLAIR | Manual | Custom U-net, ResNet152 | Internally validated only | 1p/19q: 92.15%* |
| Zhu et al [44] | 539 | IDH | 2, 3, 4 | UCSF-PDGM, In-house dataset: Sanger sequencing | In-house (single center), public (UCSF-PDGM) | T1, T1CE, T2, T2-FLAIR, ADC, SWI | Ensemble deep learning model | Radiomics | Both internally and externally validated | IDH: 88.7% |
| Aliotta et al [29] | 41 | IDH, 1p/19q | 2, 3 | IDH: IHC ± DNA pyrosequencing; 1p/19q: FISH and chromosomal microarray analysis (OncoScan) | In-house (single center) | T1, T1CE, T2, T2-FLAIR, DTI | DeepMedic | Radiomics | Internally validated only | IDH: 90%, 1p/19q: 94% |

**Table 1** continued

| Study | Total no. pts | Genes | Grade | Reference standard | Dataset(s) | MRI | Segmentation | Feature extraction | Validation | AUC |
|---|---|---|---|---|---|---|---|---|---|---|
| Alom et al [148] | 78 | IDH, 1p/19q | 2, 3, 4 | TCGA | Public (TCGA) | T1, T1CE, T2, T2-FLAIR | Radiomics (manual); DL (not undertaken, whole MRI slices) | Hybrid (VGG-Net, ResNet50, and DenseNet, Radiomics) | Internally validated only | IDH: 80.73%, 1p/19q: 87.21% * |
| González et al [61] | 99 | IDH1, 1p19q | 2 | TCGA | Public (TCGA) | T1, T1CE, T2, T2-FLAIR | Not undertaken (whole MRI slices) | Inception v3 | Internally validated only | NR |
| Riahi Samani et al [39] | 275 | IDH | 4 | NR | In-house (single center) | T1, T1CE, T2, T2-FLAIR, DTI | DeepMedic | CNNs | Internally validated only | NR |
| Sun et al [149] | 424 | IDH, 1p/19q | 2, 3, 4 | UCSF, In-house dataset: NR | In-house (single center), public (UCSF) | T1CE, T2-FLAIR | U-Net | Radiomics | Internally validated only | 96% |
| Zhao et al [41] | 202 | IDH | 2, 3, 4 | TCGA, In-house dataset NR | In-house (single center), public (TCGA) | T1CE, T2, T2-FLAIR, DWI | Not undertaken (whole MRI slices) | UCNet/F-UCNet | Both internally and externally validated | IDH: 95.63%* |
| Zhang et al [75] | 466 | IDH, 1p/19q | 2, 3, 4 | TCGA, In-house dataset NR | In-house (single center), public (TCGA) | T1CE, T2-FLAIR | Not undertaken (cropped tumor-bearing regions) | Vision Transformer with cross-attention, DenseNet-121 | Both internally and externally validated | IDH: 92%, 1p/19q: 99% |
| Yogananda et al [58] | 368 | 1p/19q | 2, 3, 4 | TCGA | Public (TCGA) | T2 | IDH network | 3D Dense-UNet | Internally validated only | 1p/19q: 95% |
| Akkus et al [90] | 159 | 1p/19q | 2, 3 | FISH | In-house (single center) | T1CE, T2 | Semi-automatic | CNNs | Internally validated only | 1p/19q: 87.7%* |
| Cao et al [85] | 954 | IDH, 1p/19q | 2, 3, 4 | TCGA, EGD; In-house dataset: IHC, DNA sequencing | In-house (multi-center), public (TCGA, EGD) | T1CE, T2-FLAIR | U-Net | Hybrid (ResNet, Graph Convolutional Network) | Both internally and externally validated | IDH: 85%, 1p/19q: 81% |

**Table 1** continued

| Study | Total no. pts | Genes | Grade | Reference standard | Dataset(s) | MRI | Segmentation | Feature extraction | Validation | AUC |
|---|---|---|---|---|---|---|---|---|---|---|
| Hu et al [150] | 256 | IDH | 2, 3, 4 | PCR amplification, Sanger sequencing | In-house (single center) | T1, T1CE, T2 | Manual | DenseNet | Internally validated only | IDH: 90.4% |
| Farahani et al [71] | 1705 | IDH, 1p/19q | 2, 3, 4 | TCGA, UCSF-PDGM, Ivy GAP, LGG-1p/19q, RHUH-GBM, UPenn-GBM, EGD | Public (TCGA, UCSF-PDGM, Ivy GAP, LGG-1p/19q, RHUH-GBM, UPenn-GBM, EGD) | T1, T1CE, T2, T2-FLAIR | MTS-UNET | MTS-UNET (SWIN-UNETR backbone) | Both internally and externally validated | IDH: 90.58%, 1p/19q: 69.22% |
| Niu et al [68] | 1185 | IDH | 2, 3, 4 | TCGA, In-house dataset: Sanger sequencing | In-house (single center), public (TCGA) | T1CE, T2-FLAIR | Manual | Hybrid (Vision transformer, Radiomics) | Externally validated only | IDH: 85.9% |
| Wu et al [30] | 2776 | IDH, 1p/19q | 2, 3, 4 | TCGA, UCSF, EGD; In-house dataset: IDH: Sanger sequencing; 1p19q: FISH | In-house (multi-center), public (TCGA, UCSF, EGD) | T1CE, T2-FLAIR | mmFormer | 3D ResNet-10 | Both internally and externally validated | IDH: 85.6%, 1p/19q: 78.8% |
| Chen et al [94] | 1806 | IDH, 1p/19q | 2, 3, 4 | BraTS 2020, EGD, LGG-1p/19q, UCSF-PDGM, REMBRANDT; In-house: NR | In-house (single center), public (BraTS 2020, EGD, LGG-1p/19q, UCSF-PDGM, REMBRANDT) | T1CE, T2 | Not undertaken (whole MRI volume + whole-brain mask) | CMTLNet | Both internally and externally validated | IDH: 86.8%, 1p/19q: 74.4% |

*Total no. pts* total number of patients, *NR* not reported, *FISH* fluorescence in situ hybridization, *IDH* isocitate dehydrogenase, *DL* deep learning, *AE* autoencoder, *CNN* convolutional neural network, *GNN* graph neural network, *RNN* recurrent neural network, *U-Net* U-Net framework, *RNN* recurrent neural network, *V-Net* volumetric neural network, *T1* T1-weighted imaging, *T1CE* T1-weighted contrast-enhanced imaging, *T1-GD* T1-weighted contrast-enhanced imaging with gadolinium, *T2-FLAIR* T2-weighted fluid-attenuated inversion recovery imaging, *DWI* diffusion-weighted imaging, *2D SS-direction HARDI* 2D SS-direction high angular resolution diffusion imaging, *DSC* dynamic susceptibility contrast MR perfusion, *DSC-PWI* dynamic susceptibility-contrast perfusion-weighted imaging, *CEST* chemical-exchange saturation transfer, *MRAGE* magnetization-prepared rapid acquisition gradient-echo, *qBOLD* quantitative blood-oxygen-level-dependent imaging, *SWI* susceptibility-weighted imaging, *TR-SPACE* T2 sampling perfection with application-optimized contrasts using different flip-angle evolutions, *AUC* area under the curve, *TCGA* the Cancer Genome Atlas, *Ivy GAP* Ivy Glioblastoma Atlas Project, *RHUH-GBM* Rio Hortega University Hospital Glioblastoma Dataset, *UPenn-GBM* University of Pennsylvania Glioblastoma Dataset, *UCSF-PDGM* University of California San Francisco Preoperative Diffuse Glioma MR, *EGD* Erasmus Glioma Database, *LGG-1p19qDeletion* LGG-1p19q deletion dataset
* AUC was not available; accuracy is reported instead

studies that did not fine-tune the segmentation models incorporated expert review and manual correction of the automatically generated ROIs to ensure accuracy [29, 35, 44, 81, 96, 100]. Additionally, clinical parameters, mainly age and sex, were incorporated in 23% of the studies [12, 28, 31, 33, 42, 43, 51, 52, 59, 63, 67, 70, 96, 101–105] (Supplementary Table 5). Regarding model development and evaluation, 37 studies performed external validation, while 65 studies relied solely on internal validation. Figure 2F summarizes the internal validation strategies, highlighting the predominance of the held-out test set approach, followed by K-fold cross-validation.

### Quality assessment
The median RQS was 15 (41.67%), ranging from 7 (19.44%) to 22 (61.11%) out of 36. In Domain 1 (mean score: 2.47 ± 0.92), most studies reported image protocols, but none included multiple time points or phantom studies; however, 71 studies conducted multiple segmentations. Domain 2 scored the highest (mean score: 5.65 ± 1.60), with 35% of studies validating their findings on external datasets. In Domain 3 (mean score: 2.60 ± 0.76), 29% of studies included multivariable analyses incorporating non-radiomic features, and 22% explored biological correlates. Domain 4 had a mean score of 2.74 ± 0.61, with most studies conducting statistical analysis. More than half of the studies used resampling techniques, though only three reported calibration statistics. All studies were retrospective and lacked prospective validation or cost-effectiveness analysis. For Domain 6 (average score: 1.51 ± 1.12), 64% of the studies used open-source data, but only 22% made their code available (Fig. 2H and Supplementary Section 4).

According to the QUADAS-2, the overall risk of bias was high in 48 studies and low in 53 studies, mainly due to limited segmentation methods or the lack of resampling techniques to mitigate overfitting. Additionally, 55 studies raised applicability concerns primarily due to a lack of validation on external datasets (Fig. 2I, J and Supplementary Section 3).

### Publication bias and statistical power
Funnel plot asymmetry and Egger's test indicated potential publication bias in IDH studies for both internal validation and test sets ($p < 0.05$), whereas no significant bias was detected in 1p/19q studies ($p > 0.05$). To account for the potential bias in IDH prediction, we applied the Trim and Fill method by Duval and Tweedie to adjust the pooled estimates of sensitivity and specificity (Supplementary Sections 7 and 9). The statistical power analysis revealed a high detection capability for larger effect sizes in most included studies but relatively lower power for detecting smaller sensitivity and specificity measures ($< 0.3$) in some studies [49, 87, 106, 107] (Supplementary Section 11).

### IDH mutation
Most models primarily targeted IDH mutation, either alone in 60% or alongside 1p/19q prediction in 36% of the studies. Over 60% of the studies focused on Grades 2, 3, and 4 gliomas. Grade 4 gliomas were exclusively studied in 12% of the experiments, while Grade 2 gliomas were addressed in only four studies. In the meta-analysis, 75% of studies used DL-based features and 25% relied on conventional radiomics; among the latter, 10 studies applied DL solely for tumor segmentation and 3 for classification.

### Meta-analysis
In both the internal validation and test cohorts, there was no significant correlation between sensitivity and specificity for IDH prediction, as indicated by SCC of 0.04 (95% CI: −0.27 to 0.33) for sensitivity and 0.01 (95% CI: −0.26 to 0.28) for specificity. In the test cohorts, the bivariate model estimated a pooled sensitivity of 80.4% (95% CI: 77.5–83.0%) and specificity of 84.6% (95% CI: 81.1–87.5%), with 95% prediction intervals ranging from 0.62 to 0.92 for sensitivity and 0.55 to 0.96 for specificity. Although unadjusted heterogeneity was moderate ($I^2 = 38.1$–69.4%, $p < 0.001$), it was markedly reduced to 2.9–3.5% after adjusting for sample size. Similar performance was achieved for internal validation cohorts (Table 2 and Supplementary Figs. 13 and 15). These results are illustrated in the SROC curves (Fig. 4A, B), which demonstrate strong overall diagnostic performance, with an area under the curve (AUC) of 0.88 for test cohorts and 0.93 for internal validation cohorts. Separate analyses of studies employing DL-based and conventional radiomic features are provided in Table 2. Additionally, forest plots with pooled estimates, including original and imputed studies using the Duval & Tweedie Trim-and-Fill method, are presented in Supplementary Section 9. Sensitivity analyses are reported in Supplementary Section 8.

### Subgroup analysis
We restricted the subgroup analysis to test cohorts (Table 3). Except for the segmentation method and the level of DL integration within the radiomics pipeline, none of the between-group differences reached statistical significance. Semi-automatic segmentation yielded the highest sensitivity, followed by DL-based and manual approaches. End-to-end DL pipelines outperformed those using DL only for feature extraction.

**Table 2** Diagnostic performance of MRI-DL models for predicting IDH mutation and 1p/19q codeletion, stratified by studies using DL for feature extraction, radiomic analysis, and all studies combined

| | Gene | Dataset | No. of studies | No. of patients | Sensitivity (95% CI) | Specificity (95% CI) | AUC | $I^2$ (Holling unadj.) | $I^2$ (Holling adj.) | p-value (SEN) | p-value (SPE) |
|---|---|---|---|---|---|---|---|---|---|---|---|
| DL | IDH | InV | 39 | 7467 | 0.86 (0.83–0.88) | 0.89 (0.87–0.91) | 0.93 | 48.7–74.9% | 2.8–4.0% | $1.18 \times 10^{-14}$ | $< 2 \times 10^{-16}$ |
| | | Test | 39 | 7284 | 0.80 (0.76–0.83) | 0.86 (0.82–0.89) | 0.89 | 49.2–77.3% | 3.5–4.4% | $< 2 \times 10^{-16}$ | $< 2 \times 10^{-16}$ |
| | 1p/19q | InV | 14 | 2695 | 0.83 (0.74–0.89) | 0.89 (0.85–0.92) | 0.93 | 60.8–82.4% | 4.2–8.0% | 3.36e13 | 2.88e10 |
| | | Test | 13 | 1599 | 0.77 (0.66–0.85) | 0.84 (0.76–0.89) | 0.87 | 49.9–72.2% | 5.0–6.0% | $3.42 \times 10^{-8}$ | $< 2 \times 10^{-16}$ |
| Radiomic | IDH | InV | 8 | 1452 | 0.85 (0.79–0.89) | 0.77 (0.69–0.84) | 0.88 | 30.4–55.7% | 1.3–1.7% | 0.010 | 4.02e-4 |
| | | Test | 13 | 959 | 0.79 (0.73–0.83) | 0.80 (0.71–0.86) | 0.82 | 0–0% | 0–0% | 0.86 | $1.14 \times 10^{-8}$ |
| | 1p/19q | InV | 0 | 0 | – | – | – | – | – | – | – |
| | | Test† | 2 | 16 | 0–83% | 73.3–84% | 0.71–0.94 | – | – | – | – |
| All studies | IDH | InV | 43† | 8133 | 0.86 (0.86–0.90) | 0.88 (0.86–0.90) | 0.93 | 52.5–74.0% | 2.9–3.9% | 6.92e-15 | <2e-16 |
| | | Test | 52 | 8243 | 0.80 (0.77–0.83) | 0.85 (0.81–0.87) | 0.88 | 38.1–69.4% | 2.9–3.5% | 3.27e11 | <2e-16 |
| | 1p/19q | InV | 14 | 2695 | 0.83 (0.74–0.89) | 0.89 (0.85–0.92) | 0.93 | 60.8–82.4% | 4.2–8.0% | 3.36e13 | 2.88e-10 |
| | | Test | 15 | 1615 | 0.75 (0.65–0.82) | 0.82 (0.75–0.88) | 0.85 | 45.9–67.5% | 5.0–5.8% | 3.18e-07 | <2e-16 |

For each subgroup, the table reports the number of studies, total patients, pooled sensitivity and specificity (95% CI), AUC, heterogeneity ($I^2$ for Holling unadjusted and adjusted), and p-value for both internal validation and test datasets

No. of studies number of studies, No. of patients number of patients, IDH isocitrate dehydrogenase, CI confidence interval, DL deep learning, AUC area under the curve, InV internal validation, SEN sensitivity, SPE specificity, unadj. unadjusted, adj. adjusted

† For 1p/19q codeletion studies in the test cohorts, only two studies [29, 106] reported radiomics results; therefore, their respective validation ranges are presented instead of pooled estimates

† Some studies included both DL and radiomics models. These were analyzed separately in the DL and Radiomics categories, but only the DL models were included in the All studies analysis

## 1p/19q Co-deletion

Approximately 5% of the studies focused only on 1p/19q co-deletion, whereas 34% addressed both 1p/19q co-deletion and IDH prediction, mainly in Grades 2 and 3 gliomas. The diagnostic performance of 1p/19q co-deletion in the internal validation and test cohorts (Fig. 3C, D) showed no significant correlation between sensitivity and specificity, with SCCs of 0.08 (95% CI: −0.47 to 0.58) and 0.03 (95% CI: −0.49 to 0.54), respectively. Meta-analysis of test datasets yielded a pooled sensitivity of 74.6% (95% CI: 64.9–82.3%) and specificity of 82.2% (95% CI: 74.8–87.8%) across fourteen experiments. Significant heterogeneity was observed ($I^2 = 45.9$–67.5%, $p < 0.001$), as shown in the SROC curves (Fig. 4C, D) by the wide, non-overlapping 95% confidence and prediction regions. However, heterogeneity was notably reduced to 5.0–5.8% following sample-size adjustment using the Holling method. Internal validation cohorts demonstrated higher predictive performance (Table 2 and Supplementary Section 9). One study used conventional radiomic features by employing DL solely for image segmentation [29]. Furthermore, sensitivity analyses are detailed in Supplementary Section 8.

## Subgroup analysis

Due to the limited number of studies per subgroup, meta-regression was not feasible for most covariates. As detailed in Table 4, studies using only in-house datasets demonstrated higher sensitivity but lower specificity compared to those trained and validated on a combination of in-house and public datasets, though the differences were not statistically significant.

## Discussion

Our systematic review and meta-analysis critically evaluated the diagnostic performance of MRI-based DL models for predicting IDH mutation and 1p/19q co-deletion in glioma patients. Consistent with prior research [17, 108–110], our findings demonstrate promising overall model performance but reveal substantial between-study heterogeneity. Notably, heterogeneity declined markedly after adjusting for sample size, indicating that most of the observed variability in sensitivity and specificity stems from sampling error rather than systematic study differences. Subsequent subgroup analyses confirmed the stability of our pooled estimates, as most examined covariates had no significant effect on model performance. Moreover, our statistical power analysis shows that while some studies had low power for small changes, most were sufficiently powered to detect pooled estimates.

Our meta-regression analysis highlighted tumor segmentation as a major source of variability. Since most

**Table 3**  Meta-regression subgroup analysis exploring heterogeneity in IDH mutation prediction within test cohorts

| Covariates | Subgroup | No. of studies | No. of patients | Sensitivity (95% CI) | *p*-value | Specificity (95% CI) | *p*-value |
|---|---|---|---|---|---|---|---|
| Glioma grade | LGG | 7 | 563 | 0.86 [0.78; 0.91] | 0.24 | 0.79 [0.71; 0.86] | 0.39 |
| | HGG | 5 | 344 | 0.73 [0.49; 0.88] | | 0.82 [0.71; 0.90] | |
| | LGG & HGG | 43 | 6620 | 0.79 [0.75; 0.83] | | 0.85 [0.80; 0.89] | |
| Clinical information | Included | 16 | 3136 | 0.79 [0.71; 0.84] | 0.66 | 0.85 [0.77; 0.90] | 0.90 |
| | Not included | 55 | 6819 | 0.80 [0.77; 0.83] | | 0.84 [0.81; 0.87] | |
| Data augmentation | Included | 39 | 7139 | 0.79 [0.76; 0.82] | 0.71 | 0.85 [0.82; 0.88] | 0.60 |
| | Not included | 32 | 2816 | 0.80 [0.74; 0.85] | | 0.83 [0.77; 0.88] | |
| Dataset | In-house | 22 | 1545 | 0.79 [0.72; 0.85] | 0.08 | 0.81 [0.75; 0.86] | 0.37 |
| | Public | 10 | 7680 | 0.86 [0.81; 0.90] | | 0.87 [0.68; 0.96] | |
| | In-house + Public | 39 | 730 | 0.79 [0.75; 0.82] | | 0.85 [0.82; 0.88] | |
| Segmentation method | DL | 12 | 4334 | 0.80 [0.76; 0.84] | 0.04 | 0.89 [0.84; 0.92] | 0.25 |
| | Manual | 8 | 1093 | 0.77 [0.67; 0.85] | | 0.83 [0.78; 0.87] | |
| | Not undertaken | 7 | 2084 | 0.72 [0.65; 0.77] | | 0.84 [0.79; 0.89] | |
| | Semi-automatic | 4 | 584 | 0.85 [0.74; 0.93] | | 0.86 [0.81; 0.90] | |
| Feature extraction | DL | 52 | 8554 | 0.80 [0.77; 0.83] | 0.93 | 0.85 [0.82; 0.88] | 0.43 |
| | Radiomics | 16 | 1097 | 0.79 [0.69; 0.86] | | 0.79 [0.68; 0.87] | |
| | DL + Radiomics | 3 | 304 | 0.79 [0.71; 0.85] | | 0.85 [0.71; 0.93] | |
| DL models | CNN | 34 | 5909 | 0.79 [0.75; 0.83] | 0.93 | 0.85 [0.82; 0.88] | 0.68 |
| | GNN | 4 | 371 | 0.80 [0.71; 0.87] | | 0.86 [0.81; 0.89] | |
| | Transformer | 7 | 1409 | 0.78 [0.70; 0.85] | | 0.84 [0.80; 0.87] | |
| DL Integration | End-to-end | 29 | 5639 | 0.81 [0.77; 0.85] | 0.37 | 0.88 [0.83; 0.92] | 0.03 |
| | Feature extraction | 26 | 3219 | 0.78 [0.74; 0.83] | | 0.82 [0.80; 0.84] | |
| MRI technique | Conventional | 62 | 9441 | 0.79 [0.77; 0.83] | 0.71 | 0.85 [0.82; 0.87] | 0.42 |
| | Advanced | 3 | 108 | 0.73 [0.20; 0.97] | | 0.73 [0.09; 0.99] | |
| | Advanced + Conventional | 6 | 406 | 0.83 [0.74; 0.90] | | 0.80 [0.71; 0.87] | |
| Validation method | Internally Validated Only | 35 | 2433 | 0.80 [0.77; 0.83] | 0.42 | 0.83 [0.80; 0.86] | 0.52 |
| | Both Internally and Externally Validated | 35 | 7339 | 0.78 [0.74; 0.82] | | 0.85 [0.80; 0.88] | |

The table reports pooled sensitivity and specificity (95% CI) for each subgroup, alongside *p*-values for between-group differences across covariates
*No. of studies* number of studies, *No. of patients* number of patients, *IDH* isocitrate dehydrogenase, *CI* confidence interval, *DL* deep learning, *AUC* area under the curve, *CNNs* convolutional neural networks, *GNN* graph neural network, *HGG* high-grade glioma, *LGG* low-grade glioma features are extracted from defined ROIs, variations in segmentation methods can significantly impact feature reproducibility [111, 112]. We also refined our QUADAS-2 assessment to include an evaluation of segmentation methods, identifying one-third of the studies as unclear or high risk due to inadequate segmentation approaches. To mitigate this, future studies should standardize and streamline the segmentation process. Utilizing robust automated segmentation tools or well-validated semi-automated pipelines can reduce inter-observer variability. Where manual segmentation is unavoidable, having multiple raters and using consensus or average segmentations might improve reliability [113]. Furthermore, several strategies have been proposed to enhance automatic segmentation. For example, applying small dilations and erosions to masks during model training can improve tolerance to boundary shifts [114]. Uncertainty in segmentation can also be estimated using ensemble methods or Monte Carlo dropout [115]. Finally, segmentation-free DL approaches offer a way to bypass manual ROI delineation entirely, potentially avoiding this source of heterogeneity. Adopting these strategies can enhance the reliability of model performance, independent of the segmentation method used.

Studies employing DL in an end-to-end approach outperformed those using DL solely for feature extraction in radiomics workflows. This direct method minimizes potential errors, enhances reproducibility, and improves predictive accuracy. Previous studies have indicated that DL, particularly CNNs, bypasses traditional complexities associated with radiomics workflows, leading to more robust feature extraction [28, 116]. However, our analysis did not reveal any significant differences in predictive performance between radiomic and DL-based features. Importantly, no variation in

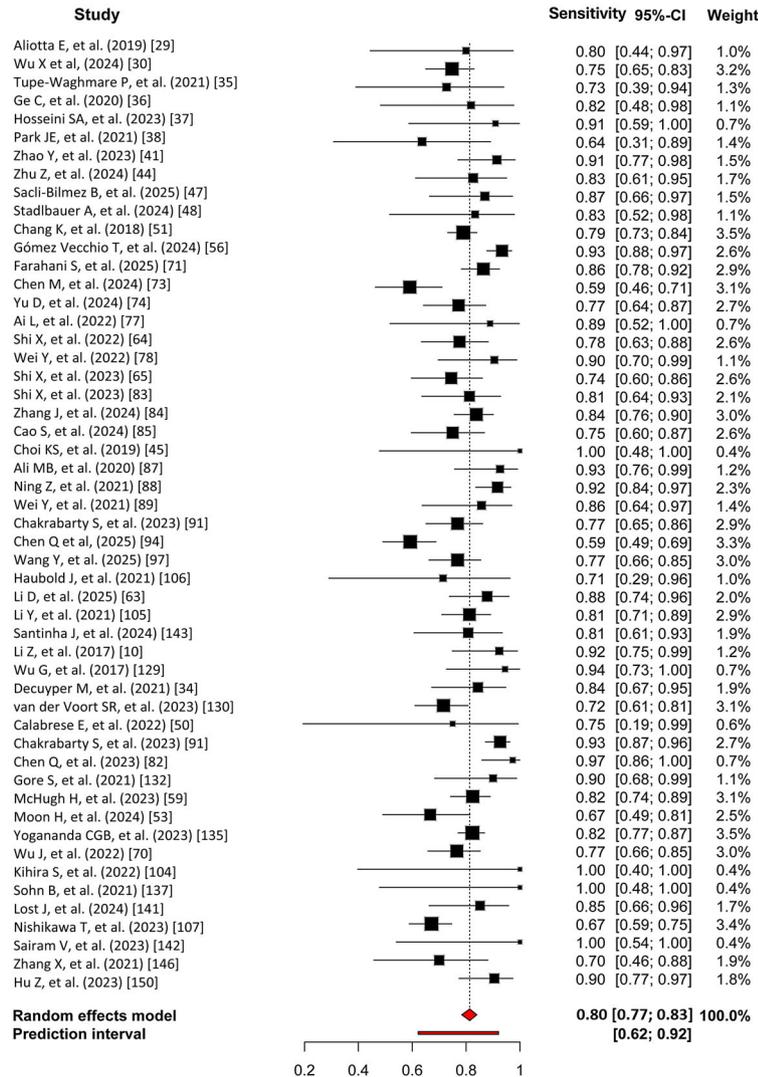

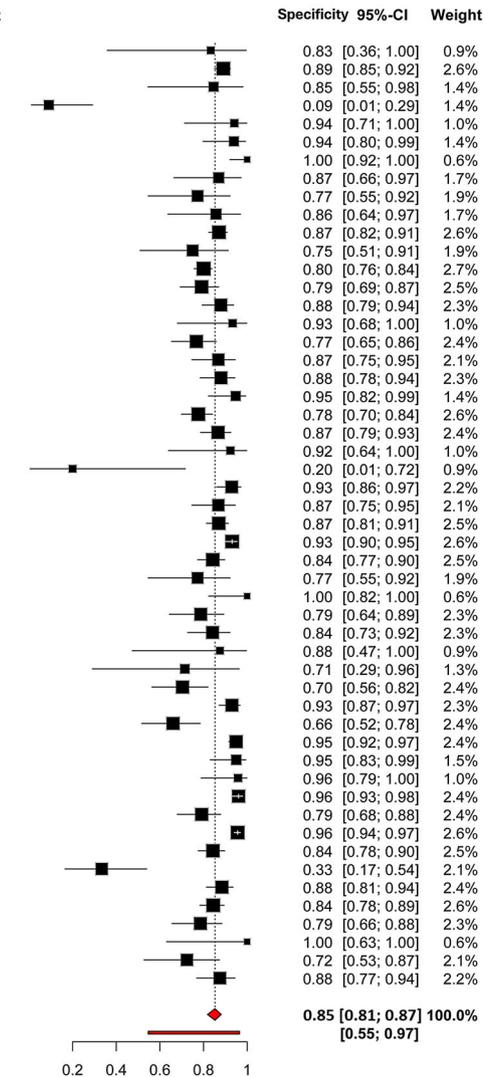

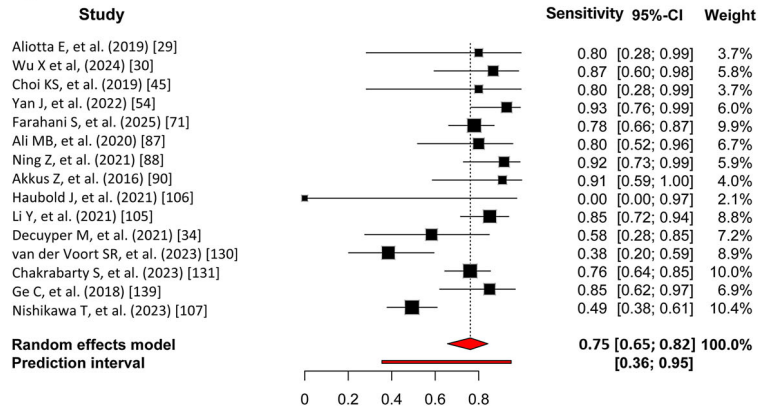

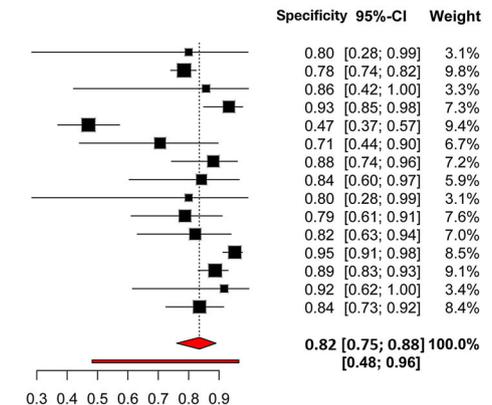

**Fig. 3** (See legend on next page.)

(see figure on previous page)

**Fig. 3** Random forest visualization of test cohorts for molecular marker prediction. **A** Sensitivity for IDH prediction. **B** Specificity for IDH prediction. **C** Sensitivity for 1p/19q prediction. **D** Specificity for 1p/19q prediction. Each plot shows the sensitivity and specificity with 95% CIs and weights for each study. The pooled estimates and prediction intervals under a random effects model are depicted at the bottom of the plots. The numbers represent pooled estimates with 95% CIs in brackets, depicted by horizontal lines. IDH, isocitrate dehydrogenase; CI, confidence interval

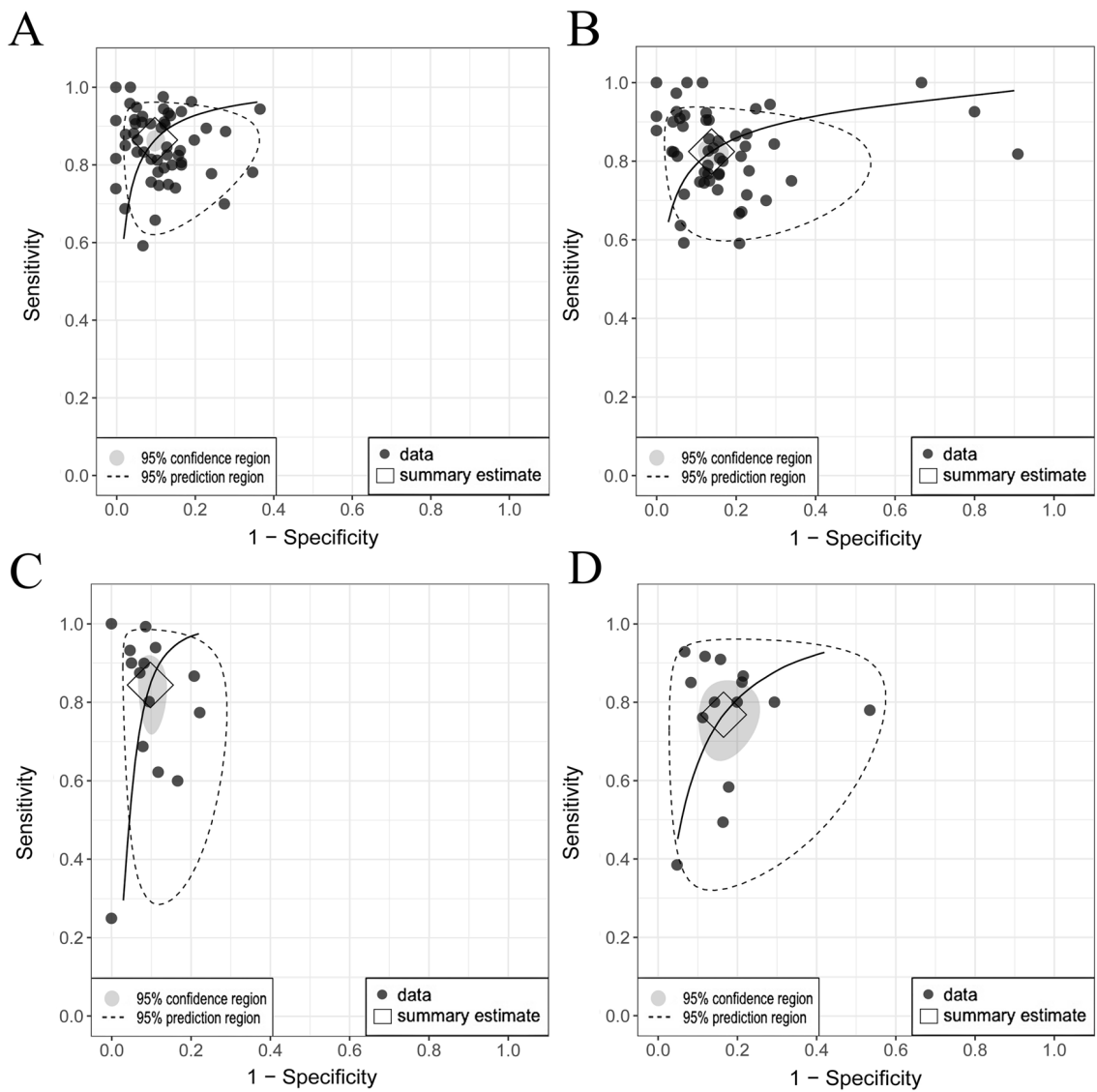

**Fig. 4** Comparison of SROC curves [26] for IDH and 1p/19q prediction in the internal validation and test cohorts. **A** IDH internal validation: pooled sensitivity 0.86 (95% CI: 0.83–0.88), specificity 0.88 (95% CI: 0.86–0.90), AUC 0.93. **B** IDH test: pooled sensitivity 0.80 (95% CI: 0.77–0.83), specificity 0.85 (95% CI: 0.81–0.87), AUC 0.88. **C** 1p/19q internal validation: pooled sensitivity 0.83 (95% CI: 0.74–0.89), specificity 0.89 (95% CI: 0.85–0.92), AUC 0.93. **D** 1p/19q test: pooled sensitivity 0.75 (95% CI: 0.65–0.82), specificity 0.82 (95% CI: 0.75–0.88), AUC 0.85. Considerable differences between the 95% confidence and prediction regions, particularly for 1p/19q codeletion, highlight significant between-study heterogeneity. SROC, summary receiver operating characteristic; IDH, isocitrate dehydrogenase; AUC, area under the curve; CI, confidence interval

performance was observed across different DL architectures, including CNNs, GNNs, and transformers. It is worth noting, though, that the limited sample size in some subgroups, such as GNN-based studies with only 304 cases, may affect the reliability of these findings.

Although not statistically significant, consistent trends were observed across both IDH and 1p/19q co-deletion studies regarding data sources. Models trained and validated on the same dataset demonstrated higher pooled sensitivity compared to those using multi-center datasets.

**Table 4** Subgroup meta-regression analysis of heterogeneity in 1p/19q codeletion prediction within test cohorts

| Covariates | Subgroup | No. of studies | No. of patients | Sensitivity (95% CI) | *p*-value | Specificity (95% CI) | *p*-value |
|---|---|---|---|---|---|---|---|
| Data augmentation | Included | 10 | 1038 | 0.79 [0.68; 0.88] | 0.31 | 0.81 [0.71; 0.88] | 0.23 |
| | Not included | 5 | 577 | 0.68 [0.43; 0.85] | | 0.88 [0.78; 0.94] | |
| Dataset | In-house | 7 | 202 | 0.83 [0.74; 0.89] | 0.24 | 0.80 [0.70; 0.87] | 0.08 |
| | In-house + public | 7 | 1245 | 0.73 [0.53; 0.86] | | 0.88 [0.82; 0.92] | |
| Validation method | Internally validated only | 7 | 202 | 0.83 [0.74; 0.89] | 0.10 | 0.80 [0.70; 0.87] | 0.51 |
| | Both Internally and externally validated | 7 | 1347 | 0.70 [0.53; 0.83] | | 0.84 [0.71; 0.92] | |

Pooled sensitivity and specificity (95% CI) are reported for each subgroup, with *p*-values denoting between-group differences for each covariate
*No. of studies* number of studies, *No. of patients* number of patients, *CI* confidence interval

In-house datasets, with standardized imaging protocols, typically offer more uniform data quality. In contrast, multi-center datasets introduce greater diversity in scanner vendors, MRI protocols, and patient populations, which can challenge models and lead to apparent performance drops, but ultimately confer more robustness. To reduce scanner- or site-specific biases, various harmonization methods have been developed [117]. At the feature level, techniques like ComBat help align radiomic feature distributions across different scanners by correcting for batch-related effects [118]. On the image level, DL−based approaches such as cycle-consistent generative adversarial networks (CycleGANs) and style transfer can be used to standardize image appearance across datasets [117, 119]. Additionally, fundamental preprocessing steps, such as correcting for bias field inhomogeneity, applying noise reduction filters, and normalizing intensities through methods like z-score scaling or histogram matching, may reduce image heterogeneity at its source [120].

The quality assessments in our systematic review revealed several areas for improvement and current limitations in the field. Consistent with previous reviews [116, 121], the median RQS score of 15 (41.67%) indicates moderate methodological quality, with deficiencies across several domains. Many studies detailed image protocols but lacked multiple time points or phantom studies, reducing reproducibility. Although over 70% of studies evaluated their models on unseen data, nearly half did not use external test datasets. This raises concerns about real-world applicability, as reflected in the RQS and QUADAS-2 assessments. Recent Food and Drug Administration (FDA) guidance on AI-enabled devices highlights that models can inadvertently overfit to features unique to a particular scanner or site [122]. To address this, it is essential to include multi-center training and external validation, and when performance declines across datasets, strategies such as domain adaptation should be employed. Foundation-based DL models offer a promising way forward. Pretrained on large, multi-institutional datasets, they tend to capture more stable and biologically meaningful features, making them more robust to variations in input data [123]. Moreover, to promote fairness across varied populations, models should undergo rigorous testing on diverse subgroups during the development and validation phases. This need is underscored by the fact that fewer than 4% of FDA-approved AI devices report race or ethnicity data [124].

Demonstrating technical performance is only one step; prospective clinical validation under real-world conditions is indispensable to bridging the gap to clinical adoption. In our review, all studies were retrospective. Prospective validation through real-time studies or clinical trials is crucial to show that DL models not only achieve high diagnostic accuracy but also improve patient outcomes compared to standard care. Unlike retrospective studies, prospective validation captures the full clinical workflow—data acquisition, model inference, and clinician decision-making—without hindsight bias, providing a more realistic assessment of the model [125]. Prospective validation also builds the case for regulatory approval and clinical acceptance, as required by related standards such as International Organization for Standardization (ISO) 13485 [126] and International Electrotechnical Commission (IEC) 62304 [127]. For instance, ISO 13485 involves risk management, documentation of design processes, and predefined acceptance criteria for performance. Collaborating with clinical partners to test the model prospectively in a workflow-simulated environment can generate the clinical evidence needed for eventual translation. Finally, incorporating DL into clinical workflows demands compatibility with electronic health records, clinician training, and robust IT infrastructure to support continuous model updates and real-time data integration. It incurs hardware, software, staffing, and maintenance costs that hospitals must weigh against potential benefits [128]. Overcoming these challenges is essential to move DL models from research into practice and advancing personalized oncology care.

This systematic review has several limitations. We focused on top-performing DL models and categorized them broadly due to a scarcity of articles. Nevertheless, we considered variations such as including clinical data, radiomic features, or different MRI sequences within a single study as separate experiments for more detailed analysis. However, these findings are observational rather than causal because randomization did not occur between studies, which is typical in most meta-analyses [22]. There may be other confounding variables influencing these results. Although reconstructing 2 × 2 tables increased the number of studies eligible for meta-analysis, imputation may introduce minor biases. Moreover, we did not assess potential patient overlap across studies. Approximately 26% of included studies relied exclusively on public datasets (mainly TCIA). While this raises the possibility of patient-level overlap, excluding these studies could introduce bias, as model performance on the same dataset can vary considerably depending on the DL framework. Our subgroup analysis further confirmed that the segmentation method and DL integration, rather than dataset origin, were the primary sources of heterogeneity.

In conclusion, our review highlights the promising performance of MRI-based DL models in accurately predicting IDH and 1p/19q co-deletion in glioma patients. To enhance the rigor and facilitate clinical translation of DL models for glioma molecular diagnosis, we propose the following minimum standards identified by our comprehensive analysis: use validated automated or consensus-based segmentation protocols, harmonize multi-center MRI data through methods such as ComBat or DL-based style transfer, incorporate phantom studies to assess feature stability, perform independent external validations without model retraining, and open data and code sharing. The next critical steps are to embed these models in prospective, multi-institutional clinical trials, integrating them into electronic health record workflows, assessing diagnostic accuracy, clinical impact, and cost-effectiveness in real time, and gathering the regulatory evidence needed for safe and effective routine use in neuro-oncology.

## Abbreviation

| | |
|---|---|
| AUC | Area under the curve |
| CAE | Convolutional autoencoder |
| CI | Confidence interval |
| CNN | Convolutional neural network |
| CycleGAN | Cycle-consistent generative adversarial network |
| DL | Deep learning |
| DSC | Dynamic susceptibility contrast |
| FDA | Food and Drug Administration |
| GNN | Graph neural network |
| IDH | Isocitrate dehydrogenase |
| ISO | International Organization for Standardization |
| RNN | Recurrent neural network |
| ROC | Receiver operating characteristic |
| ROI | Region of interest |
| RQS | Radiomics quality score |
| SCC | Spearman correlation coefficient |
| SROC | Summary receiver operating characteristic |
| TCIA | The cancer imaging archive |

## Supplementary information
The online version contains supplementary material available at https://doi.org/10.1007/s00330-025-11898-2.

## Acknowledgements
We would like to thank Dr Mary Simons for her valuable assistance in developing the search strategy.

## Author contributions
Somayeh Farahani: conceived and designed the analysis, collected the data, contributed data or analysis tools, performed the analysis, and wrote the paper. Marjaneh Hejazi: conceived and designed the analysis, and other contributions: reviewed the article before submission for both spelling and grammar, as well as intellectual content. Mehnaz Tabassum: collected the data and performed the analysis: RQS analysis. Antonio Di Ieva: conceived and designed the analysis, and other contributions: reviewed the article before submission for both spelling and grammar, as well as intellectual content. Neda Mahdavifar: performed the analysis: QUADAS-2 tool, and performed the analysis. Sidong Liu: conceived and designed the analysis, and other contributions: organized and supervised the course of the project, and took overall responsibility for the article. Additionally, reviewed the article before submission for both spelling and grammar, as well as intellectual content.

## Funding
This work was partially supported by an NHMRC Ideas Grant (GNT202035). Open Access funding enabled and organized by CAUL and its Member Institutions.

## Compliance with ethical standards

### Guarantor
The scientific guarantor of this publication is Siding Liu.

### Conflict of interest
The authors of this manuscript declare no relationships with any companies, whose products or services may be related to the subject matter of the article.

### Statistics and biometry
One of the authors, Neda Mahdavifar, holds a PhD in epidemiology and biostatistics and has substantial expertise in statistical analysis.

### Informed consent
This study includes a systematic literature review and meta-analysis following the Preferred Reporting Items for Systematic Reviews and Meta-analysis (PRISMA) guidelines. Ethical approval was unnecessary due to the nature of the study.

### Ethical approval
This study includes a systematic literature review and meta-analysis following the Preferred Reporting Items for Systematic Reviews and Meta-analysis (PRISMA) guidelines. Ethical approval was unnecessary due to the nature of the study. This study is registered on PROSPERO, number CRD42024542505.

### Declaration of generative AI and AI-assisted technologies in the writing process
During the preparation of this work, the author used Grammarly to improve language and readability. After using this tool, the authors reviewed and edited the content as needed and take full responsibility for the content of the publication.

**Study subjects or cohorts overlap**
NA.

**Methodology**

- Retrospective
- Diagnostic or prognostic study
- Multicenter study

**Author details**
¹Department of Medical Physics and Biomedical Engineering, School of Medicine, Tehran University of Medical Sciences, Tehran, Iran. ²Centre for Health Informatics, Australian Institute of Health Innovation, Macquarie University, Sydney, NSW, Australia. ³Computational NeuroSurgery (CNS) Lab, Faculty of Medicine, Health and Human Sciences, Macquarie Medical School, Macquarie University, Sydney, NSW, Australia. ⁴Department of Epidemiology & Biostatistics, School of Public Health, Tehran University of Medical Sciences, Tehran, Iran.

## Publisher's Note